\documentclass[10pt,journal,twocolumn]{IEEEtran}
\IEEEoverridecommandlockouts
\usepackage{cite}
\usepackage{amsmath,amssymb,amsfonts,booktabs}
\usepackage{algorithmic}
\usepackage{graphicx}
\usepackage{textcomp}
\usepackage{xcolor}
\usepackage{enumerate}
\usepackage{epstopdf}
\usepackage{subcaption}
\usepackage{amsthm}

\usepackage{array}
\definecolor{mygray}{RGB}{240,240,240}

\usepackage{algorithm}
\usepackage{algorithmic}
\allowdisplaybreaks[4]

\def\BibTeX{{\rm B\kern-.05em{\sc i\kern-.025em b}\kern-.08em
		T\kern-.1667em\lower.7ex\hbox{E}\kern-.125emX}}
\begin{document}

\title{\huge {Latency Minimization for Multi-AAV-Enabled ISCC Systems with Movable Antenna} }

\author{
\IEEEauthorblockN{  {
Yiyang Chen\IEEEauthorrefmark{1}\IEEEauthorrefmark{2},
Wenchao Liu\IEEEauthorrefmark{1},
Chunjie Wang\IEEEauthorrefmark{2}\IEEEauthorrefmark{3},
Yinyu Wu\IEEEauthorrefmark{4}\IEEEauthorrefmark{2},
Xuhui Zhang\IEEEauthorrefmark{4},
and Yanyan Shen\IEEEauthorrefmark{2}} \\ }
\vspace{3.5pt}
\IEEEauthorblockA{ \small
\IEEEauthorrefmark{1}Southern University of Science and Technology, Shenzhen, China\\
\IEEEauthorrefmark{2}Shenzhen Institutes of Advanced Technology, Chinese Academy of Sciences, Shenzhen, China\\
\IEEEauthorrefmark{3}University of Chinese Academy of Sciences, Beijing, China\\
\IEEEauthorrefmark{4}Shenzhen Future Network of Intelligence Institute, The Chinese University of Hong Kong, Shenzhen, China\\
\vspace{-0.75em} 
}
\vspace{-18.5pt}
}

\maketitle

\begin{abstract}
This paper investigates an autonomous aerial vehicle (AAV)-enabled integrated sensing, communication, and computation system, with a particular focus on integrating movable antennas (MAs) into the system for enhancing overall system performance.
Specifically, multiple MA-enabled AVVs perform sensing tasks and simultaneously transmit the generated computational tasks to the base station for processing.
To minimize the maximum latency under the sensing and resource constraints, we formulate an optimization problem that jointly coordinates the position of the MAs, the computation resource allocation, and the transmit beamforming.
Due to the non-convexity of the objective function and strong coupling among variables, we propose a two-layer iterative algorithm leveraging particle swarm optimization and convex optimization to address it.
The simulation results demonstrate that the proposed scheme achieves significant latency improvements compared to the baseline schemes.
\end{abstract}


\section{Introduction}

Integrated sensing and communication (ISAC) has emerged as a promising paradigm that seamlessly integrates radar sensing and wireless communication functionalities into a unified system, enabling efficient spectrum utilization and hardware resource sharing \cite{10012421}.
Although ISAC offers significant potential for future intelligent networks, its practical deployment faces several challenges, among which signal blockage caused by buildings and other terrestrial obstacles severely limits system performance.
To address this issue, the integration of autonomous aerial vehicles (AAVs), also known as unmanned aerial vehicles or drones, into ISAC systems has attracted increasing attention, as the AAV-enabled ISAC platforms can exploit line-of-sight (LoS) propagation to bypass ground obstructions, thereby enhancing overall system reliability and quality of service.
Several recent studies \cite{10100680, 10879807} have investigated the feasibility and benefits of the AAV-enabled ISAC in several application scenarios.
The authors in \cite{10100680} proposed a joint beamforming and AAV trajectory design to improve the system throughput.
The authors in \cite{10879807} considered the antenna configuration and receiver type, aiming to enhance the communication performance of the AAV while meeting the sensing requirements.

It is worth noting that in ISAC systems, the sensing tasks often generate a massive amount of data that must be rapidly processed and analyzed to enable real-time decision-making and accurate environmental modeling.
However, due to their limited on-board computing capabilities, AAVs struggle to handle computational intensive tasks, such as urban map reconstruction, which typically requires the aggregation and processing of data from multiple AAVs, necessitating the offloading of such tasks to ground base stations (BSs) for centralized computation.
This leads to the emergence of integrated sensing, communication, and computation (ISCC) systems, which not only facilitate efficient sensing data acquisition and low-latency transmission, but also enhance system intelligence and resource utilization through joint orchestration of communication and computation resources.
Recent works \cite{10233771, 10462480, 10757511} have highlighted the potential of the ISCC frameworks in addressing challenges from different scenarios.
The authors in \cite{10233771} proposed an AAV-enabled ISCC system and demonstrated its performance advantages by minimizing the age of information of all users.
The authors in \cite{10462480} considered an ISCC framework that incorporates non-orthogonal multiple access to mitigate interference and enhance throughout performance.
The authors in \cite{10757511} investigated a mobile edge computing (MEC)-assisted ISAC system and designed a radar signal processing method to minimize energy consumption of devices.

Recently, movable antenna (MA) technology has attracted increasing research interest as a novel approach that enables dynamic adjustment of antenna positions within a limited physical space to optimize signal propagation \cite{liu2025movable}.
Owing to its inherent spatial degrees of freedom, MA offers significant potential to enhance key system performance metrics, such as signal quality, channel capacity, interference coordination, and energy efficiency \cite{10318061}.
Previous works \cite{10620306, 10654366, zhang2025movable} have explored the deployment applications of MAs in various wireless communication scenarios.
The authors in \cite{10620306} proposed a wireless-powered MEC system with MAs, introduced three MA placement configurations, and maximized the total computation rate.
The authors in \cite{10654366, zhang2025movable} studied the downlink and uplink transmission in an AAV-enabled communication system equipped with the MA, respectively, aiming to maximize the data rate of the users.

Despite the emerging interest in AAV-enabled ISCC systems, existing studies remain limited, particularly in scenarios where sensing-generated data is treated as computational tasks to be offloaded for centralized processing.
Integrating MAs into such systems, leveraging their spatial flexibility to further enhance overall performance, remains unexplored, which motivates our work.
In this paper, we investigate an MA-assisted multi-AAV-enabled ISCC system, where the AAVs equipped with MAs perform target sensing and simultaneously offload the resulting computational tasks to a ground BS for processing.
Our goal is to minimize the maximum processing latency of all sensing tasks by jointly optimizing the position of the MA array, the AAV transmit beamforming, and the BS computation resource allocation.
To address this problem, we propose an efficient two-layer iterative algorithm based on particle swarm optimization (PSO) and convex optimization.
Simulation results demonstrate that the proposed scheme achieves significant latency improvements compared to the benchmark schemes.

% \textit{Organizations:} The rest of this paper is organized as follows.
% Section II introduces the MA-assisted multi-AAV-enabled ISCC system and formulates the problem of minimizing the maximum processing latency of sensing tasks.
% Section III proposes a two-layer iterative algorithm and analyzes the convergence and complexity of the proposed algorithm.
% Section IV shows the numerical results of the proposed scheme versus multiple benchmark schemes.
% Finally, Section V concludes the paper.

\textit{Notations:} The notations are introduced below.
$ {{\mathbb{C}}^{M\times N}} $ denotes the $ M \times N $ complex matrix $ \mathbb{C} $.
$ \mathrm{j} $ represents the imaginary unit, where $ {\mathrm{j}^2} = -1 $.
For a generic matrix $ {\boldsymbol{G}} $, $ {{\boldsymbol{G}}^{\mathsf{H}}} $, $ {{\boldsymbol{G}}^{\mathsf{T}}} $, $ \mathsf{tr}({\boldsymbol{G}}) $, and $\mathsf{rank}({\boldsymbol{G}})$ denote the conjugate transpose, transpose, trace, and rank of $ {\boldsymbol{G}} $, respectively.
$\| \boldsymbol{a} \|$ denotes the norm of $ {\boldsymbol{a}} $.

\section{System Model and Problem Formulation}

\begin{figure}[!htbp]
	\centering
	\includegraphics[width=0.8\linewidth]{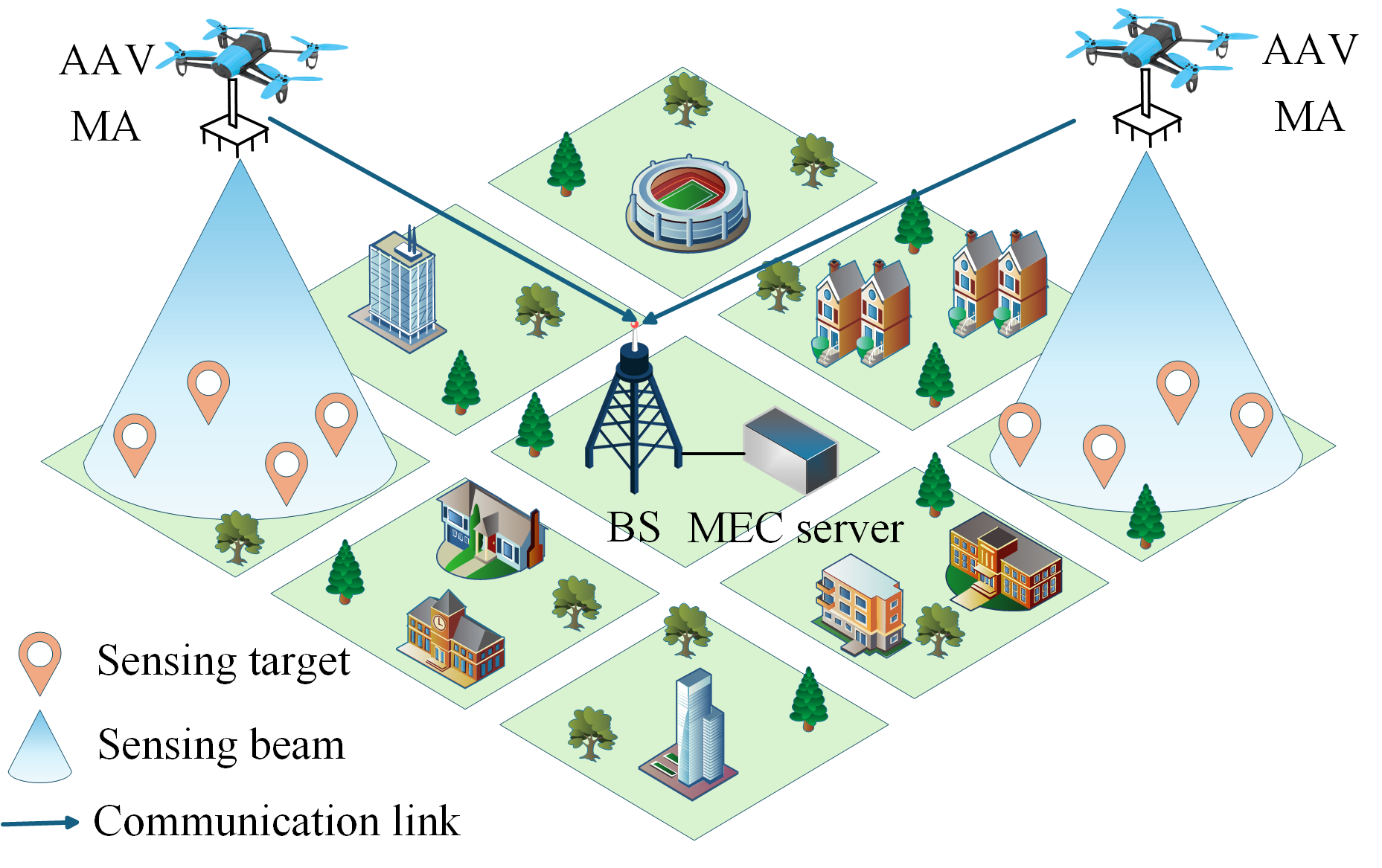}
	\caption{The MA-assisted AAV-enabled ISCC system.}
	\label{fig:MA_AAV_ISCC}
    \vspace{-2em} % 可选：根据需要调整这里以减少上下空间
\end{figure}

\subsection{System Model}

As illustrated in Fig. \ref{fig:MA_AAV_ISCC}, we consider an MA-assisted AAV-enabled ISCC system, which consists of a ground BS and $M$ AAVs equipped with MAs.
The set of AAVs is denoted as $\mathcal{M}= \{1, 2, \cdots, M\}$.
In the ISCC system, there exist numerous sensing targets, which typically exhibit a clustered distribution due to their inherent sensing characteristics, e.g., sudden regional wildfires or area-wide urban map reconstruction tasks, thereby forming distinct sensing target areas.
To mitigate the adverse effects caused by terrain factors and building blockages, one AAV is dispatched for serving each sensing target area \cite{10980172}.
Considering a more general scenario, the AAVs are assumed to hover in the air while performing sensing tasks.
Such sensing tasks generate data that require rapid processing, analysis, and aggregation. However, due to their inherent limitations in computation resources, the AAVs are unable to handle these computation-intensive tasks efficiently.
Fortunately, the AAVs can offload these tasks to the BS for centralized computation. Specifically, the BS is equipped with an MEC server, which is capable of meeting the stringent latency requirements of such tasks.
To address the aforementioned sensing and processing issues, the MA equipped on the AAV simultaneously senses the target area and transmits the generated computational tasks to the BS.
To mitigate transmission interference, frequency division multiple access (FDMA) is employed, where different AAVs utilize the orthogonal frequency bands for communicating with their served sensing areas and the BS.

We assume that each two-dimension MA array deployed on different AAVs is configured identically and optimized individually. It is assumed that each MA consists of $N$ antennas, and the set of the antennas is given by $\mathcal{N} = \{1, 2, \cdots, N\}$.
Moreover, the position of the MA array on the $m$-th AAV is denoted by $\boldsymbol{u}_{m} = [\boldsymbol{u}_{m,1}, \boldsymbol{u}_{m,2}, \cdots, \boldsymbol{u}_{m,N}]^{\mathsf{T}}$, where $\boldsymbol{u}_{m,n} = [x_{m,n},y_{m,n}]$ represents the position of the $n$-th MA on the $m$-th AAV.
Specifically, $x_{m,n}$ and $y_{m,n}$ are expressed in units of wavelength $\lambda$.
It is further assumed that the antenna positions are subject to physical constraints.
Specifically, each antenna can only be adjusted within a bounded range, i.e., $\boldsymbol{u}_{m,n} \in \mathcal{R}_u$, where $\mathcal{R}_u$ is the maximum allowable square region with size $U \times U$.
Additionally, to mitigate mutual coupling effects between adjacent elements, a minimum separation distance $u_{\min}$ must be maintained between any pair of antennas, i.e., $\| \boldsymbol{u}_{m,i} - \boldsymbol{u}_{m,j} \| \geq u_{\min}$, where $\{ i,j \} \in \mathcal{N}$, and $\ i \neq j$.

Since each AAV is responsible for both sensing the corresponding target area and transmitting the generated task information to the BS in a simultaneous manner, the signal transmitted by the $m$-th AAV can be expressed as
\begin{equation}
    \boldsymbol{z}_{m}^{\mathrm{Tx}} = \boldsymbol{w}_{m} s_{m} + \boldsymbol{v}_{m},
\end{equation}
where $\boldsymbol{w}_{m} \in \mathbb{C}^{N \times 1}$ denotes the downlink information beamforming vector directed toward the BS, $s_{m} \in \mathbb{C}$ represents the data symbol intended for the BS with unit power, i.e., $\mathbb{E} \{ {| s_{m} |}^{2} \} = 1$, and $\boldsymbol{v}_{m} \in \mathbb{C}^{N \times 1}$ corresponds to the dedicated sensing signal transmitted by the $m$-th AAV. The covariance matrix of $\boldsymbol{v}_{m}$ is defined as $\boldsymbol{V}_{m} \triangleq \mathbb{E} \{ \boldsymbol{v}_{m} \boldsymbol{v}_{m}^\mathsf{H} \}$.
In addition to the transmit requirement, the total transmit power consumed by the $m$-th AAV must satisfy the constraint ${\| \boldsymbol{w}_{m} \|}^{2} + \mathsf{tr}(\boldsymbol{V}_{m}) \leq P_{m}^{\max}$, where $P_{m}^{\max}$ denotes the maximum available transmit power at the $m$-th AAV.

Specifically, considering the sensing model, the steering vector corresponding to the MA array mounted on the $m$-th AAV and directed towards its associated sensing target area can be represented as
\begin{equation}
    \boldsymbol{g}_{m} = \left [ e^{{\mathrm{j}} \frac{2\pi}{\lambda} \rho_{m,1}}, 
    e^{{\mathrm{j}} \frac{2\pi}{\lambda} \rho_{m,2}},
    \cdots, 
    e^{{\mathrm{j}} \frac{2\pi}{\lambda} \rho_{m,N}} \right ]^\mathsf{T},
\end{equation}
where $\rho_{m,n}$ can be given by
\begin{equation}
    \rho_{m,n} = x_{m,n} \sin{\theta_m} \cos{\vartheta_m} + y_{m,n} \sin{\theta_m} \sin{\vartheta_m},
\end{equation}
with $\theta_m$ and $\vartheta_m$ representing the angle of departure (AoD) from the $m$-th AAV to the target area of interest in the vertical and horizontal planes, respectively \cite{8974403}.
To quantify the sensing capability of the AAV towards its designated target area, we employ the beampattern gain in that direction as the sensing performance metric.
To improve the sensing effectiveness, both the information-carrying signal and the dedicated sensing signal are jointly utilized for dedicated sensing \cite{9916163, 10086626}.
Based on this design, it is required that the beampattern gain of the $m$-th AAV towards the target area satisfies a minimum performance criterion, i.e., $\boldsymbol{g}_{m}^\mathsf{H} ( \boldsymbol{w}_{m} \boldsymbol{w}_{m}^\mathsf{H} + \boldsymbol{V}_{m} ) \boldsymbol{g}_{m} \geq d_{m}^2 \Gamma_{\min}$, where $d_{m}$ denotes the distance between the $m$-th AAV and its associated target area, and $\Gamma_{\min}$ represents the minimum required sensing performance threshold.

Accordingly, each AAV is deemed to have successfully completed its sensing mission and generated the corresponding computation tasks for further processing only if it meets the prescribed sensing performance requirements.
The computation tasks generated by each AAV comprise two distinct types: radar-based sensing and onboard camera imaging \cite{10966042, 10729736}.
The total amount of computation tasks generated by the $m$-th AAV is denoted by $D_{m}$, which is proportional to the volume of raw data collected during the sensing phase and reflects the computational burden associated with subsequent processing.

We consider an air-to-ground (A2G) channel model between the AAVs and the BS that incorporates both large-scale and small-scale fading effects. 
Specifically, the A2G channel gain between the BS and the $m$-th AAV is given by
\begin{equation}
    \boldsymbol{h}_{m} = \sqrt{\frac{h_0}{d_{\mathrm{BS},m}^2}} \left( \sqrt{\frac{\kappa}{\kappa + 1}} \boldsymbol{a}_{m} + \sqrt{\frac{1}{\kappa + 1}} \boldsymbol{\tilde{h}}_{m} \right) ,
\end{equation}
where $h_0$ denotes the reference channel gain at a distance of $1 \mathrm{m}$, $\boldsymbol{\tilde{h}}_{m}$ represents the non-LoS (NLoS) component, which is the complex Gaussian random vector with zero mean and unit covariance matrix, $\kappa$ is the Rician factor indicating the ratio between the power of the LoS and NLoS components, $d_{\mathrm{BS},m}$ stands for the distance between the BS and the $m$-th AAV, and $\boldsymbol{a}_{m}$ denotes the steering vector corresponding to the MA array on the $m$-th AAV directed towards the BS.
More specifically, the steering vector $\boldsymbol{a}_{m}$ can be expressed as
\begin{equation}
    \boldsymbol{a}_{m} = \left [ e^{{\mathrm{j}} \frac{2\pi}{\lambda} \varrho_{m,1}}, 
    e^{{\mathrm{j}} \frac{2\pi}{\lambda} \varrho_{m,2}},
    \cdots, 
    e^{{\mathrm{j}} \frac{2\pi}{\lambda} \varrho_{m,N}} \right ]^\mathsf{T},
\end{equation}
where $\varrho_{m,n}$ can be given by
\begin{equation}
    \varrho_{m,n} = x_{m,n} \sin{\phi_m} \cos{\varphi_m} + y_{m,n} \sin{\phi_m} \sin{\varphi_m},
\end{equation}
with $\phi_m$ and $\varphi_m$ representing the AoD from the $m$-th AAV to the BS in the vertical and horizontal planes, respectively.
Hence, the signal received at the BS can be expressed as
\begin{equation}
    z^{\mathrm{Rx}} = \sum_{m=1}^M \boldsymbol{h}_{m}^\mathsf{H} \boldsymbol{w}_{m} s_{m} + n,
\end{equation}
where $n$ is the additive white Gaussian noise at the BS with variance $\sigma^2$.
Specifically, we consider that the BS is equipped with a sophisticated receiver architecture capable of mitigating the interference caused by the dedicated sensing signals \cite{10879807}.
This type of receiver can reconstruct the interference using the channel state information and the pre-known sensing signal, and subsequently subtract this interference from the received signal before decoding the desired information signal.
Moreover, considering the application of FDMA, the achievable communication rate between the $m$-th AAV and the BS can be written as
\begin{equation}
\label{R}
    R_m = \frac{B}{M} \log_2 \left ( 1 + \frac{{|\boldsymbol{h}_{m}^\mathsf{H} \boldsymbol{w}_{m}|}^2}{ \sigma^2 } \right ),
\end{equation}
where $B$ represents the channel bandwidth.
Thus, the corresponding transmission latency required for the $m$-th AAV to offload its generated tasks to the BS can be given by
\begin{equation}
    T_m^{\mathrm{tran}} = \frac{D_m}{R_m}.
\end{equation}

In the considered system, only the MEC server deployed at the BS is capable of processing the information from the sensing targets. As a result, the computation latency required for the BS to process the sensing tasks offloaded by the $m$-th AAV is given by
\begin{equation}
    T_m^{\mathrm{comp}} = \frac{\beta D_m}{f_{\mathrm{BS},m}},
\end{equation}
where $\beta$ denotes the number of CPU cycles needed to process one bit of data, and $f_{\mathrm{BS},m} \geq 0$ represents the allocated computation resource for processing the task from the $m$-th AAV at the BS.
The total computational capacity available at the BS is limited by a maximum value, denoted as $f_{\mathrm{BS}}^{\max}$. Hence, the computation resource allocation at the BS must satisfy the constraint $\sum_{m=1}^M f_{\mathrm{BS},m} \leq f_{\mathrm{BS}}^{\max}$.

Specifically, the data size of the computation results is much smaller than that of the tasks, thus the latency required for transmitting the computing results by the BS can be neglected.
Hence, the processing latency associated with the sensing tasks generated by the $m$-th AAV can be expressed as the sum of transmit and computation latency, i.e.,
\begin{equation}
    T_m = T_m^{\mathrm{tran}} + T_m^{\mathrm{comp}}.
\end{equation}

\subsection{Problem Formulation}

In this paper, we aim to minimize the maximum processing latency of the sensing tasks within the considered MA-assisted AAV-enabled ISCC system.
By jointly optimizing the position of the MA array $\boldsymbol{U} \triangleq \{ \boldsymbol{u}_{m}, \forall m \}$, the transmit beamforming at the AAV, including information beamforming $\boldsymbol{W} \triangleq \{ \boldsymbol{w}_{m}, \forall m \}$ and dedicated sensing beamforming $\boldsymbol{V} \triangleq \{ \boldsymbol{V}_{m}, \forall m \}$, and the computation resource allocation of the BS $\boldsymbol{F} \triangleq \{ f_{\mathrm{BS},m}, \forall m \}$, the maximum processing latency minimization problem can be formulated as
\begin{subequations}
	\begin{align}
\textbf{P1}:\ &\min\limits_{\{\boldsymbol{U}, \boldsymbol{W}, \boldsymbol{V}, \boldsymbol{F}\}} \ \max\limits_{m \in \mathcal{M}} \{ T_m \} \nonumber\\
{\mathrm{s.t.}}\ 
& \boldsymbol{u}_{m,n} \in \mathcal{R}_u, \ \forall m, \ \forall n, \label{p1a}\\
& \| \boldsymbol{u}_{m,i} - \boldsymbol{u}_{m,j} \| \geq u_{\min}, \ \forall m, \ \{ i,j \} \in \mathcal{N}, \ i \neq j, \label{p1b}\\
& {\| \boldsymbol{w}_{m} \|}^{2} + \mathsf{tr}(\boldsymbol{V}_{m}) \leq P_{m}^{\max}, \ \forall m, \label{p1c}\\
& \boldsymbol{g}_{m}^\mathsf{H} ( \boldsymbol{w}_{m} \boldsymbol{w}_{m}^\mathsf{H} + \boldsymbol{V}_{m} ) \boldsymbol{g}_{m} \geq d_{m}^2 \Gamma_{\min}, \ \forall m, \label{p1d}\\
& 0 \leq f_{\mathrm{BS},m}, \ \sum_{m=1}^M f_{\mathrm{BS},m} \leq f_{\mathrm{BS}}^{\max}, \ \forall m, \label{p1e}
	\end{align}
\end{subequations}
where (\ref{p1a}) and (\ref{p1b}) represent the position constraints of the MA array, (\ref{p1c}) 
denotes the AAV transmit power limitation, (\ref{p1d}) denotes the requirements of sensing beampattern gain, and (\ref{p1e}) is the computation resource allocation constraints for the BS.
The non-convexity of problem \textbf{P1} arises from the complex interactions among the variables in the constraint \eqref{p1d} and the objective function, posing significant challenges in determining the globally optimal solution. To address this issue, we employ a two-layer iterative algorithm to efficiently solve the problem.

\section{Proposed Two-Layer Iterative Algorithm}

\begin{figure}[!htbp]
	\centering
	\includegraphics[width=0.65\linewidth]{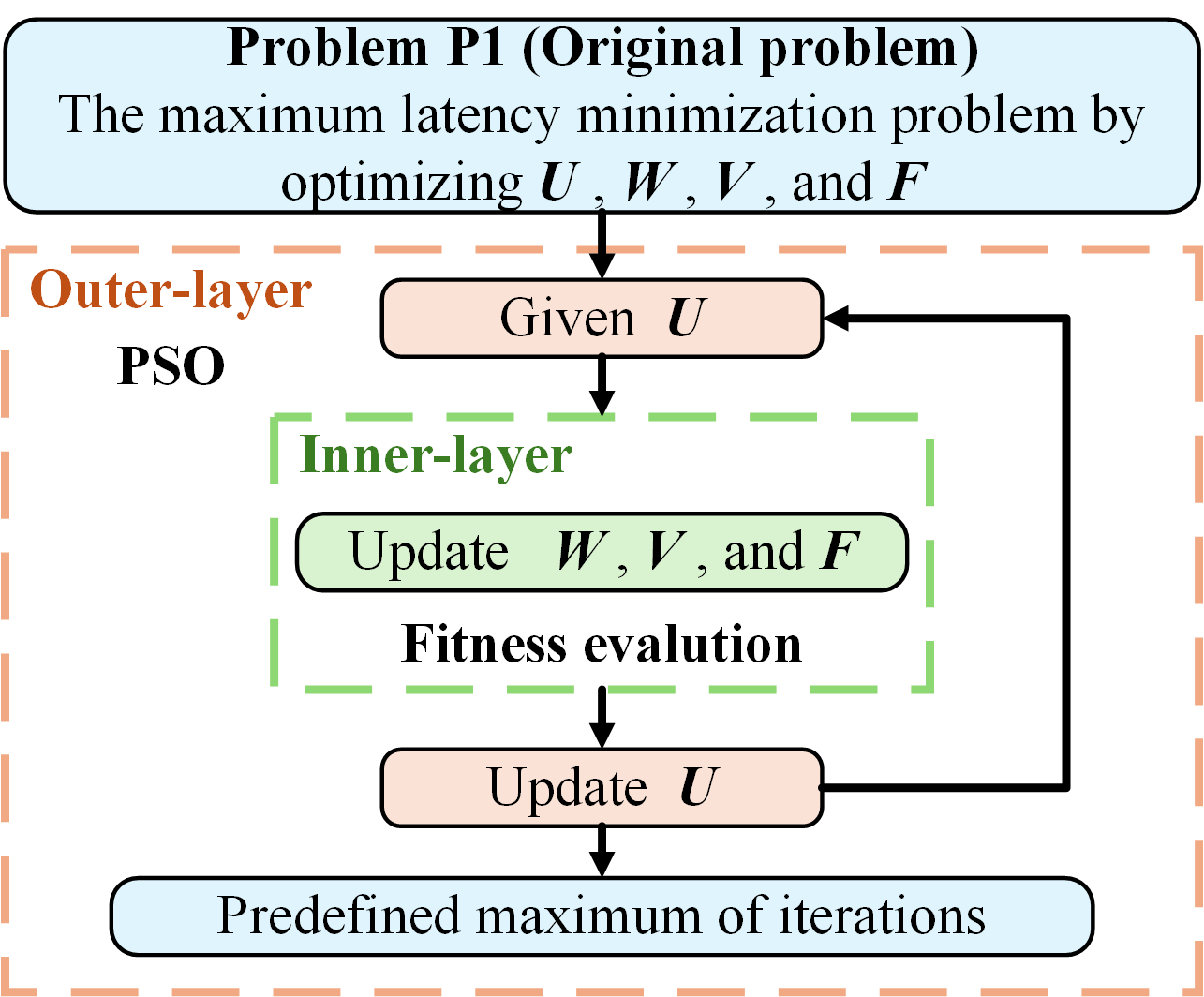}
	\caption{The MA-assisted AAV-enabled ISCC system.}
	\label{fig:MA_ISCC_flow}
    \vspace{-2em} % 可选：根据需要调整这里以减少上下空间
\end{figure}

To address the problem \textbf{P1}, we develop a two-layer iterative algorithm based on PSO and convex optimization.
In the inner-layer, given the position of the MA array, we utilize convex optimization to jointly optimize the transmit beamforming and the computation resource allocation.
In the outer-layer, we employ the PSO algorithm to optimize the position of the MA array, where the fitness value of each particle is obtained based on the optimization results in the inner-layer.
The flowchart of the proposed two-layer iterative algorithm is shown in Fig. \ref{fig:MA_ISCC_flow}.

\subsection{Transmit Beamforming and Computation Resource Allocation Optimization}

In the inner-layer of the proposed algorithm, since there is no coupling between the transmit beamforming and the computation resource allocation, they can be optimized simultaneously.
However, it is evident that the objective function of the problem \textbf{P1} is non-concave and difficult to handle.
To address this issue, we transform the original problem into an equivalent form by introducing an auxiliary variable $\Phi = \max\limits_{m \in \mathcal{M}} \{ T_m \}$.
Therefore, when $\{\boldsymbol{U}\}$ is given, the problem of optimizing the computation resource allocation of the BS $\boldsymbol{F}$ and the transmit beamforming of the AAV, including the information beamforming $\boldsymbol{W}$ and the sensing beamforming $\boldsymbol{V}$, can be reformulated as
\begin{subequations}
	\begin{align}
\textbf{P2}:\ &\min\limits_{\{\boldsymbol{W}, \boldsymbol{V}, \boldsymbol{F}, \Phi\}} \ \Phi \nonumber\\
{\mathrm{s.t.}}\ 
& \Phi \geq T_m, \ \forall m, \label{p2a}\\
&\text{(\ref{p1c}), (\ref{p1d}), and (\ref{p1e})}. \nonumber
	\end{align}
\end{subequations}

Then, we define $\boldsymbol{W}_m = \boldsymbol{w}_m \boldsymbol{w}_m^\mathsf{H}$, which follows that $\boldsymbol{W}_m \succeq 0$ and $\mathsf{rank}(\boldsymbol{W}_m) \leq 1$.
By substituting $\boldsymbol{{W}'} \triangleq \{ \boldsymbol{W}_m, \forall m \}$ into \textbf{P2}, it can be reformulated as
\begin{subequations}
	\begin{align}
\textbf{P3}:\ &\min\limits_{\{\boldsymbol{{W}'}, \boldsymbol{V}, \boldsymbol{F}, \Phi\}} \ \Phi \nonumber\\
{\mathrm{s.t.}}\ 
& \Phi \geq \frac{D_m}{R_m(\boldsymbol{W}_m)} + \frac{\beta D_m}{f_{\mathrm{BS},m}}, \ \forall m, \label{p3a}\\
& \mathsf{rank}(\boldsymbol{W}_m) \leq 1, \ \forall m, \label{p3b}\\
& \mathsf{tr}(\boldsymbol{W}_{m}) + \mathsf{tr}(\boldsymbol{V}_{m}) \leq P_{m}^{\max}, \ \forall m, \label{p3c}\\
& \boldsymbol{g}_{m}^\mathsf{H} ( \boldsymbol{W}_{m} + \boldsymbol{V}_{m} ) \boldsymbol{g}_{m} \geq d_{m}^2 \Gamma_{\min}, \ \forall m, \label{p3d}\\
&\text{(\ref{p1e})}, \nonumber
	\end{align}
\end{subequations}
where
\begin{equation}
    R_m(\boldsymbol{W}_m) = \frac{B}{M} \log_2 \left ( 1 + \frac{\mathsf{tr}(\boldsymbol{H}_{m} \boldsymbol{W}_{m})}{ \sigma^2 } \right ),
\end{equation}
with $\boldsymbol{H}_{m} \triangleq \boldsymbol{h}_m \boldsymbol{h}_m^\mathsf{H}$.

It can be clearly verified that in the problem \textbf{P3}, the function $\frac{1}{f_{\mathrm{BS},m}}$ is convex with respect to $f_{\mathrm{BS},m}$, and it can be directly solved.
Additionally, to facilitate the derivation of a feasible solution to the problem \textbf{P3}, we introduce auxiliary variables $\gamma_m$ to replace $R_m(\boldsymbol{W}_m)$, such that $0 < \gamma_m \leq R_m(\boldsymbol{W}_m)$.
Consequently, the problem \textbf{P3} can be finally reformulated as
\begin{subequations}
	\begin{align}
\textbf{P4}:\ &\min\limits_{\{\boldsymbol{{W}'}, \boldsymbol{V}, \boldsymbol{F}, \Phi, \boldsymbol{\gamma}\}} \ \Phi \nonumber\\
{\mathrm{s.t.}}\ 
& \Phi \geq \frac{D_m}{\gamma_m} + \frac{\beta D_m}{f_{\mathrm{BS},m}}, \ \forall m, \label{p4a}\\
& \gamma_m \leq R_m(\boldsymbol{W}_m), \ \forall m, \label{p4b}\\
&\text{(\ref{p1e}), (\ref{p3b}), (\ref{p3c}), and (\ref{p3d})}, \nonumber
	\end{align}
\end{subequations}
where $\boldsymbol{\gamma} \triangleq \{ \gamma_m, \forall m \}$.

In problem \textbf{P4}, the constraint \eqref{p3b} remains non-convex. To overcome this challenge, we relax it using semi-definite relaxation, thereby reformulating the problem into a standard convex optimization problem.
While the relaxed solution does not necessarily guarantee a rank-one structure, there always exists an rank-one solution $\boldsymbol{W}_m^{*}$ for the problem \textbf{P4} \cite{9916163, 10693833}.
A detailed proof is omitted here due to space limitations.
As a result, all remaining constraints and objective function in the problem \textbf{P4} are convex, which can be effectively solved using the convex optimization solver, e.g., CVX.

\subsection{MA Array Position Optimization}

In the outer-layer of the proposed two-layer
iterative algorithm, with $\{\boldsymbol{W}, \boldsymbol{V}, \boldsymbol{F}\}$ obtained in the inner-layer, the problem of optimizing the position of the MA array $\boldsymbol{U}$ can be reformulated as
\begin{subequations}
	\begin{align}
\textbf{P5}:\ &\min\limits_{\{\boldsymbol{U}\}} \ \max\limits_{m \in \mathcal{M}} \{ T_m(\boldsymbol{u}_{m}) \} \nonumber\\
{\mathrm{s.t.}}\ 
&\text{(\ref{p1a}), (\ref{p1b}) and (\ref{p1d})}. \nonumber
	\end{align}
\end{subequations}

We utilize the PSO to solve the problem \textbf{P5}.
Specifically, $P$ search solutions are randomly initialized from the feasible set as the particle swarm.
The set of solutions is denoted as $\mathcal{P}= \{1, 2, \cdots, P\}$.
We denote the solution of the $p$-th particle during the $i$-th iteration as $\boldsymbol{u}_{(i)}^{p} = [{(\boldsymbol{u}_{1,(i)}^{p})}^{\mathsf{T}}, {(\boldsymbol{u}_{2,(i)}^{p})}^{\mathsf{T}}, \cdots, {(\boldsymbol{u}_{M,(i)}^{p})}^{\mathsf{T}}]^{\mathsf{T}}$, where $\boldsymbol{u}_{m,(i)}^{p} = [u_{m,1,(i)}^{p}, u_{m,2,(i)}^{p}, \cdots, u_{m,N,(i)}^{p}]^{\mathsf{T}}$.
Then, the local optimal solution of the $p$-th particle in the $i$-th iteration is given by
\begin{equation}
    \boldsymbol{u}_{(i)}^{p*} = {\arg\min}_{\boldsymbol{u}^{p} \in \{\boldsymbol{u}_{(1)}^{p*}, \cdots, \boldsymbol{u}_{(i)}^{p*}\}} \max\limits_{m \in \mathcal{M}} \{ T_m(\boldsymbol{u}_{m}) \}.
\end{equation}
Subsequently, each particle updates the position according to individual and swarm experiences, gradually converging towards the global optimal solution.
The global optimal solution of the particle swarm in the $i$-th iteration is given by
\begin{equation}
    \boldsymbol{u}_{(i)}^{*} = {\arg\min}_{\boldsymbol{u}^{p} \in \{\boldsymbol{u}_{(i)}^{1*}, \cdots, \boldsymbol{u}_{(i)}^{P*}\}} \max\limits_{m \in \mathcal{M}} \{ T_m(\boldsymbol{u}_{m}) \}.
\end{equation}

Therefore, the exploitation of the $p$-th particle is updated by
\begin{equation}
\label{velocity}
    \nu_{(i+1)}^{p} = \omega \nu_{(i)}^{p} + c_1 \tau_1 (\boldsymbol{u}_{(i)}^{p*} - \boldsymbol{u}_{(i)}^{p}) + c_2 \tau_2 (\boldsymbol{u}_{(i)}^{*} - \boldsymbol{u}_{(i)}^{p}),
\end{equation}
where $\nu_{(i)}^{p}$ denotes the update velocity of the $p$-th particle during the $i$-th iteration, $c_1$ and $c_2$ denote the individual learning factor and global learning factor, respectively, $\tau_1$ and $\tau_2$ are random parameters uniformly distributed within $[0,1]$, and $\omega$ represents the inertia weight.
In particular, to balance the search speed and accuracy of the particle swarm, $\omega$ is continuously decreased during the iterations, which can be expressed as
\begin{equation}
\label{omega}
    \omega = \left( \omega_{\max} - \frac{(\omega_{\max}-\omega_{\min})i}{i_{\max}} \right),
\end{equation}
where $\omega_{\max}$ and $\omega_{\min}$ denote the maximum and minimum values of $\omega$, respectively, and $i_{\max}$ represents the maximum number of iterations.
Hence, the solution is updated by
\begin{equation}
\label{position}
    \boldsymbol{u}_{(i+1)}^{p} = \boldsymbol{u}_{(i)}^{p} + \varpi \nu_{(i+1)}^{p},
\end{equation}
where $\varpi$ represents a constant that controls the update.

We further consider the constraints of the problem \textbf{P5}.
For the constraint \eqref{p1a}, the position components of each particle can be projected onto the corresponding minimum and maximum bounds to ensure that the particle remains within
the feasible region throughout the optimization process.
For constraints \eqref{p1b} and \eqref{p1d}, we introduce adaptive penalty factor into the fitness function, i.e.,
\begin{equation}
\label{fitness}
    \delta(\boldsymbol{u}_{(i)}^{p}) = \max\limits_{m \in \mathcal{M}} \{ T_m(\boldsymbol{u}_{m}) \} + \psi_1 |\Psi_1(\boldsymbol{u}_{(i)}^{p})| + \psi_2 |\Psi_2(\boldsymbol{u}_{(i)}^{p})|,
\end{equation}
where $\delta(\boldsymbol{u}_{(i)}^{p})$ is the fitness function, $\Psi_1(\boldsymbol{u}_{(i)}^{p})$ and $\Psi_2(\boldsymbol{u}_{(i)}^{p})$ represent the sets of particles that violate the constraint \eqref{p1b} and \eqref{p1d}, respectively, $\psi_1$ and $\psi_2$ are penalty parameters introduced to enforce constraint satisfaction.
It is worth noting that the objective function value in Eq. (\ref{fitness}) is updated iteratively based on the $\{\boldsymbol{W}, \boldsymbol{V}, \boldsymbol{F}\}$ obtained in each inner-layer iteration.
By evaluating the fitness of each particle, the local optimal solution and the global optimal solution are iteratively updated until the algorithm converges.

\subsection{Convergence and Complexity Analysis}

The two-layer iterative algorithm for solving the problem \textbf{P1} is detailed in Algorithm 1.
The algorithm operates in a two-layer iterative manner, where the outer-layer optimizes the MA array position and the inner-layer jointly optimizes the computation resource allocation and transmit beamforming. The process continues until the predefined maximum number of iterations is reached.
As the objective function of the problem \textbf{P1} does not increase during the iterations, Algorithm \ref{Alg1} is ensured to converge \cite{10741192}.

\begin{algorithm} \small
\caption{Two-Layer Iterative Algorithm for Solving ($\textbf{P1}$)}
\label{Alg1}
\begin{algorithmic}
    \REQUIRE An initial feasible solution $\{\boldsymbol{U}, \boldsymbol{W}, \boldsymbol{V}, \boldsymbol{F}\}$;
    \STATE \textbf{Initialize:} Iteration number $i=1$, and the maximum number of iterations $i_{\max}$;
    \REPEAT
    \STATE Update the inertia weight $\omega$ by Eq. (\ref{omega});
    \STATE \textbf{Initialize:} Particle $p=1$, and the number of particles $P$;
    \REPEAT
    \STATE Update the velocity and position by Eq. (\ref{velocity}) and Eq. (\ref{position})
    \STATE Update $\boldsymbol{W}, \boldsymbol{V}, \boldsymbol{F}$ by solving the problem (\textbf{P4});
    \STATE Evaluate the fitness value by Eq. (\ref{fitness});
    \STATE Update $p = p+1$;
    \UNTIL The particle $p > P$;
    \STATE Update $\boldsymbol{U}$;
    \STATE Update $i = i+1$;
    \UNTIL The iteration number $i > i_{\max}$;
    \ENSURE The optimal solution $\{\boldsymbol{U}^{*}, \boldsymbol{W}^{*}, \boldsymbol{V}^{*}, \boldsymbol{F}^{*}\}$.
\end{algorithmic}
\end{algorithm}

The complexity is analyzed as follows.
The problem \textbf{P1} is solved using the two-layer iterative algorithm. 
The inner-layer is handled using the interior-point method via the CVX, and its computational complexity can be expressed as $\mathcal{O}(( 2MN^2 + M )^{3.5} \log(\epsilon^{-1}))$, where $\epsilon$ represents the search accuracy.
The outer-layer is solved using the PSO, hence the computational complexity of Algorithm \ref{Alg1} can be given by $\mathcal{O}(i_{\max} P ( 2MN^2 + M )^{3.5} \log(\epsilon^{-1}))$ \cite{10100680}.

\section{Numerical Results}

In this section, we present numerical results to validate the effectiveness of the proposed scheme.
The simulation considers a region with dimensions of $600\mathrm{m} \times 600\mathrm{m} \times 50\mathrm{m}$, which contains $M = 3$ AAVs with $N = 4$ MAs.
The minimum separation distance is $u_{\min} = \lambda / 2$ and the maximum allowable square region is $\mathcal{R}_u = 20\lambda \times 20\lambda$.
Each sensing target area is located at the ground beneath its corresponding AAV, and it is offset by $20\mathrm{m}$ in the horizontal direction. 
The amount of generated computation tasks satisfies $D_{m} \in [1 \times 10^7\mathrm{bits}, 1.5 \times 10^7\mathrm{bits}]$.
The reference channel gain is $h_0 = -60\mathrm{dB}$ and the noise power is $\sigma^2 = -110 \mathrm{dBm}$.
The number of CPU cycles needed to process one bit of data by the BS is $\beta = 100$.
The bandwidth is $B = 3\mathrm{MHz}$.
Other parameters of the PSO are set as follows.
The particle swarm size and the maximum number of iterations are $P = 100$ and $i_{\max} = 200$.
The individual and global learning factors are $c_1 = c_2 =1.5$.
The maximum and minimum values of the inertia weight are $\omega_{\max} = 0.9$ and $\omega_{\min} = 0.4$.
The penalty parameters are $\psi_1 = \psi_2 =100$.
For comparison purposes, the following benchmark schemes are considered:
1) \textit{Fixed-position antenna array} (FPA). This scheme assumes that the positions of antenna array are fixed.
2) \textit{Random-position antenna array} (RPA). The positions of antenna array are randomly selected while satisfying the constraints (\ref{p1a}) and (\ref{p1b}).

\begin{figure}[!htbp]
	\centering
	\includegraphics[width=0.85\linewidth]{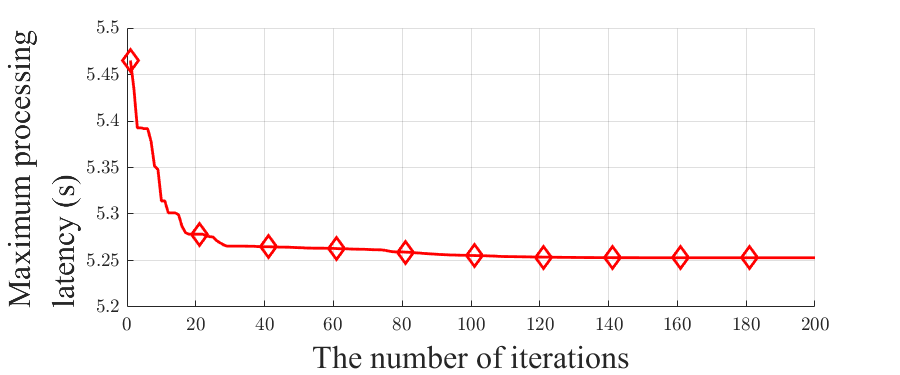}
	\caption{Maximum processing latency versus the number of iterations.}
	\label{fig:Convergence}
    \vspace{-1em} % 可选：根据需要调整这里以减少上下空间
\end{figure}

Fig.~\ref{fig:Convergence} depicts the relationship between the maximum processing latency and the number of iterations.
The simulation results indicate that our proposed scheme converges within a moderate number of iterations (e.g., 140 iterations), exhibiting a favorable convergence rate and achieving a substantial reduction in the maximum processing latency, thereby validating the efficacy of the proposed scheme.

\begin{figure}[!t]
    \centering
    \begin{minipage}[b]{0.48\columnwidth}
        \centering
        \includegraphics[width=\linewidth]{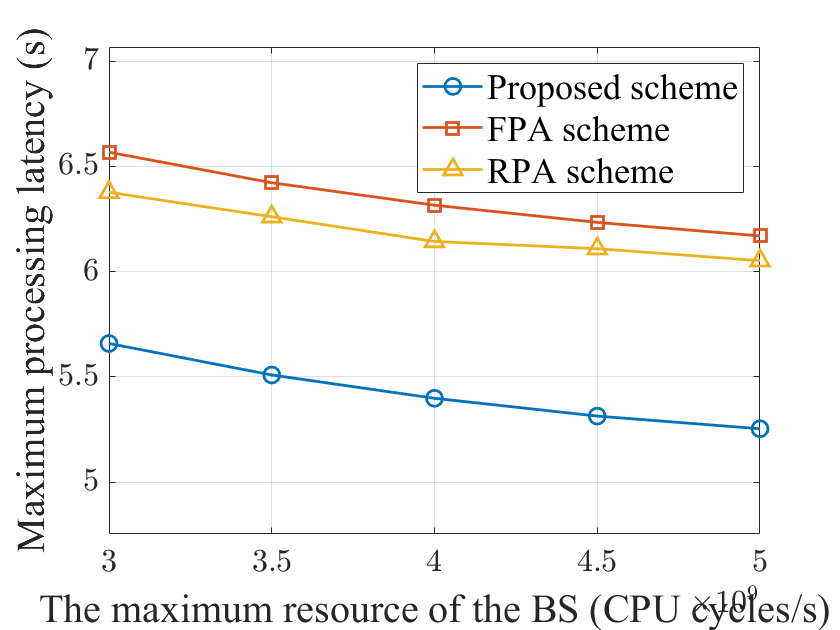}
        \caption{Maximum processing latency versus the maximum computation resource of BS.}
        \label{fig:F_T}
    \end{minipage}
    \hfill
    \begin{minipage}[b]{0.48\columnwidth}
        \centering
        \includegraphics[width=\linewidth]{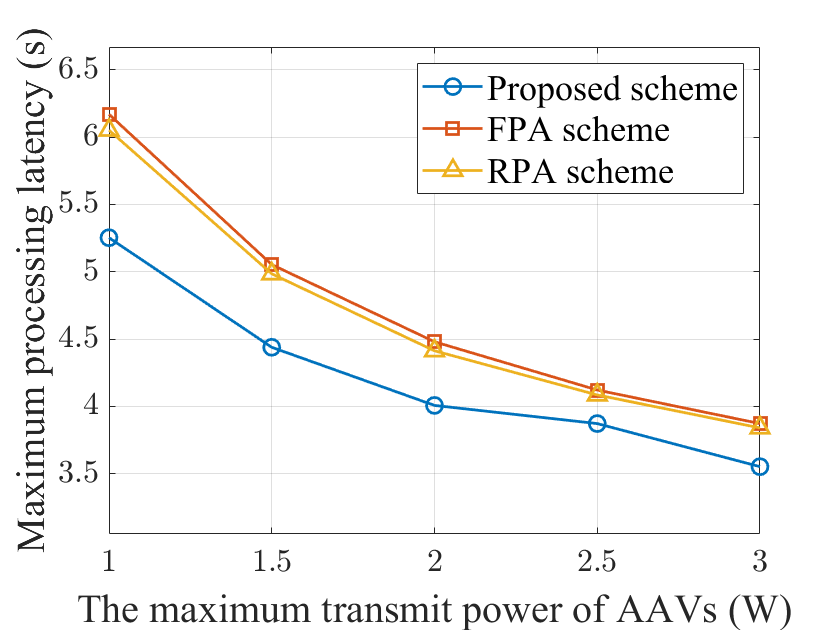}
        \caption{Maximum processing latency versus the maximum transmit power of AAVs.}
        \label{fig:P_T}
    \end{minipage}
    \vspace{-1em} % 可选：根据需要调整这里以减少上下空间
\end{figure}

Fig.~\ref{fig:F_T} and Fig.~\ref{fig:P_T} illustrate the maximum processing latency versus the maximum computation resource of the BS, and the maximum transmit power of the AAVs, respectively.
As illustrated in Fig.~\ref{fig:F_T}, the maximum processing latency decreases with an increase in the maximum available computation resources. This improvement comes from the effective allocation of computation resources, enabling the BS to selectively allocate resources to tasks from different AAVs.
In Fig.~\ref{fig:P_T}, as the maximum transmit power increases, the advantage of optimizing the transmit beamforming becomes evident, allowing for improved communication rates and consequently lower maximum processing latency.
Notably, in both figures, the proposed scheme consistently achieves better performance than the baseline schemes, demonstrating the flexibility and advantages offered by the optimized position of the MAs.

\begin{figure}[!t]
    \centering
    \begin{minipage}[b]{0.48\columnwidth}
        \centering
        \includegraphics[width=\linewidth]{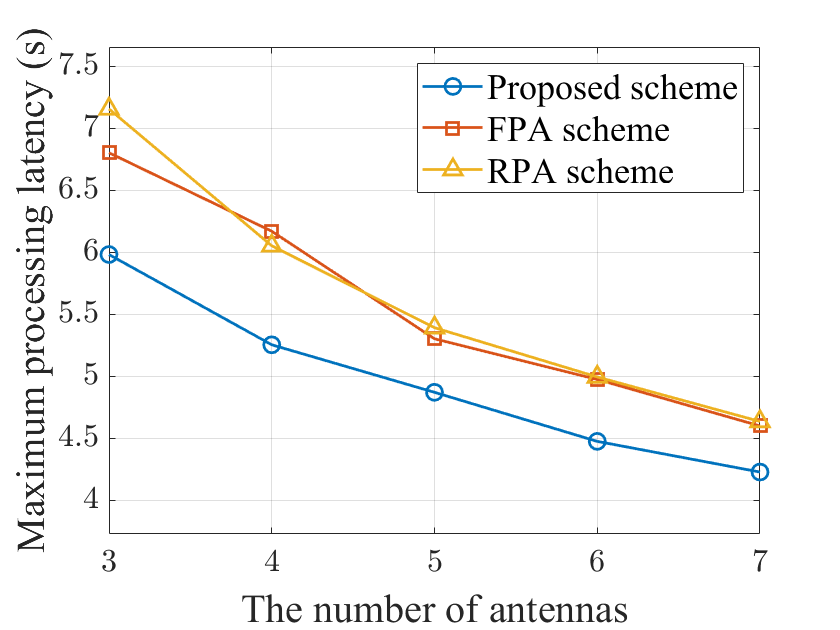}
        \caption{Maximum processing latency versus the number of antennas.}
        \label{fig:MA_T}
    \end{minipage}
    \hfill
    \begin{minipage}[b]{0.48\columnwidth}
        \centering
        \includegraphics[width=\linewidth]{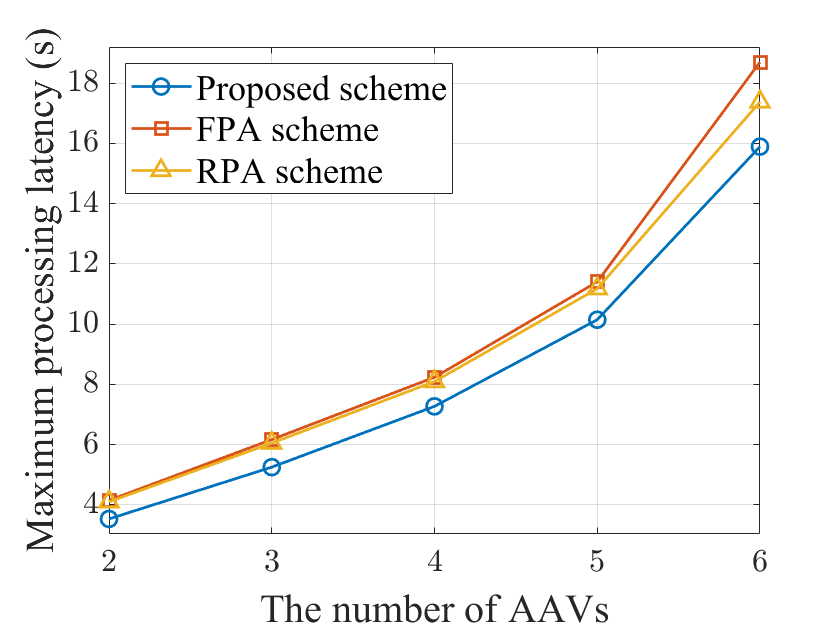}
        \caption{Maximum processing latency versus the number of AAVs.}
        \label{fig:AAV_T}
    \end{minipage}
    \vspace{-2em} % 可选：根据需要调整这里以减少上下空间
\end{figure}

Fig.~\ref{fig:MA_T} and Fig.~\ref{fig:AAV_T} show the maximum processing latency versus the number of antennas and the AAVs, respectively.
In Fig.~\ref{fig:MA_T}, as the number of antennas increases, the channel gain between the AAV and the BS improves, leading to a reduction in the maximum processing latency.
Fig.~\ref{fig:AAV_T} illustrates that as the number of the AAVs increases, the sensing areas increase and the computational tasks grow, thereby increasing the maximum processing latency.
Specifically, in both cases, the proposed scheme demonstrates significant latency improvement over the baseline schemes, which further validates the superiority of the MAs.

\section{Conclusion}

This paper investigates an MA-assisted AAV-enabled ISCC system, in which the MAs mounted on the AAVs simultaneously sense the target area and offload the generated tasks to the BS for computation.
By jointly optimizing the position of the MAs, the transmit beamforming of the AAVs, and the computation resource allocation of the BS, the maximum processing latency of all sensing tasks is minimized.
Numerical results demonstrated that the proposed scheme achieves significant performance improvement in terms of the processing latency compared to the benchmark schemes.

This paper focuses on AAVs operating in a stationary hovering mode. Future research can explore more complex AAV mobility scenarios.
In addition, more variables can be introduced to further enhance the optimization performance.

\bibliographystyle{ieeetr}
\bibliography{myref}
\end{document}